\documentclass[10pt]{article}
\usepackage{graphicx}
\usepackage{amsmath}
\usepackage{amssymb}
\usepackage{caption2}
\begin{document}
\begin{center}
{\large {\bf \sc{  Tetraquark state candidates: $Y(4260)$, $Y(4360)$, $Y(4660)$ and $Z_c(4020/4025)$  }}} \\[2mm]
Zhi-Gang  Wang \footnote{E-mail: zgwang@aliyun.com.  }     \\
 Department of Physics, North China Electric Power University, Baoding 071003, P. R. China
\end{center}

\begin{abstract}
In this article, we construct the axialvector-diquark-axialvector-antidiquark type tensor current to interpolate both the vector and axialvector tetraquark states,  then calculate the contributions of the vacuum condensates up to dimension-10  in the operator product expansion, and obtain the QCD sum rules for  both the vector and axialvector tetraquark states.  The numerical results support assigning the $Z_c(4020/4025)$ to be the $J^{PC}=1^{+-}$ diquark-antidiquark type tetraquark state, and  assigning the $Y(4660)$ to be the $J^{PC}=1^{--}$ diquark-antidiquark type tetraquark state. Furthermore, we take the $Y(4260)$ and $Y(4360)$ as the mixed charmonium-tetraquark states,  and construct the two-quark-tetraquark type tensor currents to study the masses and pole residues. The  numerical results  support assigning the $Y(4260)$ and $Y(4360)$ to be the mixed charmonium-tetraquark states.
 \end{abstract}



\section{Introduction}

In 2005, the BaBar collaboration  studied the initial-state radiation process  $e^+ e^- \to \gamma_{ISR} \pi^+\pi^- J/\psi$ and observed the $Y(4260)$   in the $\pi^+\pi^- J/\psi$ invariant-mass spectrum, the measured mass and width are  $\left(4259 \pm 8 {}^{+2}_{-6}\right) \,\rm{MeV}$ and  $\left(88 \pm 23 {}^{+6}_{-4} \right)\, \rm{MeV}$, respectively \cite{BaBar4260-0506}.
In 2007, the  Belle collaboration  studied the initial-state radiation process $e^+e^- \to \gamma_{ISR}\pi^+ \pi^- \psi^{\prime}$, and  observed two structures $Y(4360)$ and $Y(4660)$ in the $\pi^+ \pi^- \psi^{\prime}$ invariant mass distributions   at $(4361\pm 9\pm 9)\, \rm{MeV}$ with a width of \, $(74\pm 15\pm 10)\,\rm{ MeV}$ and   $(4664\pm 11\pm 5)\,\rm{ MeV}$ with a width of \, $(48\pm 15\pm 3) \,\rm{MeV}$, respectively \cite{Belle4660-0707}.
In 2008, the Belle collaboration studied  the initial-state radiation process $e^+e^- \to \gamma_{ISR} \Lambda_c^+ \Lambda_c^-$   and observed a clear peak $Y(4630)$   in the $\Lambda_c^+ \Lambda_c^-$  invariant mass distribution just above the $\Lambda_c^+ \Lambda_c^-$ threshold, and determined  the mass and width to be   $\left(4634^{+8}_{-7}{}^{+5}_{-8}\right)\,\rm{MeV}$ and $\left(92^{+40}_{-24}{}^{+10}_{-21}\right)\,\rm{MeV}$, respectively \cite{Belle4630-0807}. The $Y(4660)$ and $Y(4630)$ may be the same particle according to the uncertainties of the masses and widths.

 In 2013, the BESIII collaboration  observed
the $Z^{\pm}_c(4025)$ near the $(D^{*} \bar{D}^{*})^{\pm}$ threshold in the $\pi^\mp$ recoil mass spectrum  in the process $e^+e^- \to (D^{*} \bar{D}^{*})^{\pm} \pi^\mp$, and determined  the mass and width  $M_{Z_c^\pm(4025)}=(4026.3\pm2.6\pm3.7)\,\rm{MeV}$  and $\Gamma_{Z_c^\pm(4025)}=(24.8\pm5.6\pm7.7)\,\rm{MeV}$ \cite{BES1308}.
Furthermore, the  BESIII collaboration observed the  $Z_c^\pm(4020)$   in the $\pi^\pm h_c$ mass spectrum in the process $e^+e^- \to \pi^+\pi^- h_c$, and determined the  mass and width  $M_{Z_c^\pm(4020)}=(4022.9\pm 0.8\pm 2.7)\,\rm{MeV}$   and $\Gamma_{Z_c^\pm(4020)}=(7.9\pm 2.7\pm 2.6)\,\rm{MeV}$ \cite{BES1309}. In 2014, the BESIII collaboration  observed the  $Z_c^0(4020)$   in the $\pi^0 h_c$ mass spectrum in the process $e^+e^- \to \pi^0\pi^0 h_c$  and determined the  mass    $M_{Z_c^0(4020)}=(4023.9\pm2.2\pm3.8)\,\rm{MeV}$  \cite{BES1409}. In 2015, the BESIII collaboration  observed
the $Z^{0}_c(4025)$  in the $\pi^0$ recoil mass spectrum  in the process $e^+e^- \to (D^{*} \bar{D}^{*})^{0} \pi^0$, and determined
 the mass and width  $M_{Z_c^0(4025)}=(4025.5^{+2.0}_{-4.7}\pm3.1)\,\rm{MeV}$  and $\Gamma_{Z_c^0(4025)}=(23.0\pm 6.0\pm 1.0)\,\rm{MeV}$ \cite{BES1507}. It is natural to assign the $Z_c(4020)$ and $Z_c(4025)$ to be the same particle.

There have been several tentative assignments for the $Y(4260)$, $Y(4360)$, $Y(4660)$ and $Z_c(4020)$, such as tetraquark states, molecular states, re-scattering effects, etc, for more literatures on the $X$, $Y$, $Z$ mesons, one can consult the recent reviews \cite{XYZ-review}. In this article, we will focus on  the  scenario of tetraquark  states based on the QCD sum rules.

The diquarks $q^{T}_j C\Gamma q^{\prime}_k$ have  five  structures  in Dirac spinor space, where $C\Gamma=C\gamma_5$, $C$, $C\gamma_\mu \gamma_5$,  $C\gamma_\mu $ and $C\sigma_{\mu\nu}$ for the scalar, pseudoscalar, vector, axialvector  and  tensor diquarks, respectively.  The structures
$C\gamma_\mu $ and $C\sigma_{\mu\nu}$ are symmetric, while the structures
$C\gamma_5$, $C$ and $C\gamma_\mu \gamma_5$ are antisymmetric.
The attractive interactions of one-gluon exchange  favor  formation of
the diquarks in  color antitriplet, flavor
antitriplet and spin singlet \cite{One-gluon},
 while the favored configurations are the scalar ($C\gamma_5$) and axialvector ($C\gamma_\mu$) diquark states \cite{WangDiquark,WangLDiquark}.
  The calculations based on the QCD sum rules indicate that the  heavy-light scalar and axialvector  diquark states have almost  degenerate masses \cite{WangDiquark}.
We can construct the  diquark-antidiquark type hidden charm tetraquark states \cite{WangScalarT},
\begin{eqnarray}
&&C\gamma_5 \otimes \gamma_5C\, , \nonumber\\
&&C\gamma_\mu \otimes \gamma^\mu C\, ,
\end{eqnarray}
 the  $C\gamma_5\otimes \gamma_5 C$ type  and $C\gamma_\mu \otimes  \gamma^\mu C$  type currents couple potentially    to the lowest scalar tetraquark states with the masses about $3.82\,\rm{GeV}$ \cite{WangMPLA} and $3.85\,\rm{GeV}$ \cite{WangTetraquarkCTP}, respectively.
If the contribution of an additional P-wave  to the mass is about $0.5\,\rm{GeV}$, we can construct the vector currents
\begin{eqnarray}
&&C\gamma_\alpha \otimes \partial_\mu \gamma^\alpha C\, , \nonumber \\
&&C\gamma_5\otimes\partial_\mu \gamma_5C\, ,
\end{eqnarray}
to study the  vector tetraquark states,  the estimated masses are about $4.35\,\rm{GeV}$, which happens to be the value of the mass of the  $Y(4360)$ \cite{WangTetraquarkCTP}. In Refs.\cite{ZhangHuang-PRD,ZhangHuang-JHEP}, Zhang and Huang take the $C\gamma_5\otimes \partial_\mu \gamma_5C$ type currents to study the $Y(4360)$ and $Y(4660)$ with the QCD sum rules,  and obtain the values $M_{Y(4360)}=(4.32\pm0.20)\,\rm{GeV}$ and $M_{Y(4660)}=(4.69\pm0.36)\,\rm{GeV}$, which are consistent with the rough estimation $M_{Y(4360)}=4.35\,\rm{GeV}$.

We can also construct the
\begin{eqnarray}
&&C\otimes \gamma_\mu C\, , \nonumber\\
&&C\gamma_5 \otimes \gamma_5\gamma_\mu C\, ,
\end{eqnarray}
type currents to study the vector tetraquark states \cite{WangJPG}.
One can consult Ref.\cite{ChenZhu} for more interpolating currents for the vector tetraquark states without introducing additional P-wave.
In Ref.\cite{WangJPG}, we observe that the $C\otimes \gamma_\mu C$ type and $C\gamma_5 \otimes \gamma_5\gamma_\mu C$ type   tetraquark states  have degenerate (or slightly different) masses based on the QCD sum rules, the ground state masses of the vector tetraquark states with the symbolic quark constituent $\bar{c}c\bar{q}q$ are about $4.95\,\rm{GeV}$, which is much larger than the mass of the $Y(4660)$. In Ref.\cite{Nielsen-4260-4460}, Albuquerque and Nielsen take the $
C\gamma_5 \otimes \gamma_5\gamma_\mu C$ type current to study the $Y(4660)$ with the QCD sum rules and obtain the value $M_{Y(4660)}=4.65\,\rm{GeV}$, which is in excellent agreement with the mass of the $Y(4660)$. Although both in Ref.\cite{WangJPG} and in Ref.\cite{Nielsen-4260-4460}, the standard values of the vacuum condensates are taken, in Ref.\cite{WangJPG}, the QCD spectral densities are calculated at the energy scale $\mu=1\,\rm{GeV}$ and the value $m_c(\mu=1{\rm GeV})=1.35\,\rm{GeV}$ is taken; while in Ref.\cite{Nielsen-4260-4460}, the vacuum condensates are taken at the energy scale $\mu=1\,\rm{GeV}$ and the $\overline{MS}$ mass $m_c(m_c)=1.23\,\rm{GeV}$ is taken, the energy scales of the QCD spectral densities are not specified.
In Ref.\cite{WangEPJC-4660}, we suggest a formula $\mu=\sqrt{M^2_{X/Y/Z}-(2{\mathbb{M}}_c)^2}$   with the effective mass ${\mathbb{M}}_c$  to determine  the energy scales of the QCD spectral densities of the hidden charmed tetraquark states, and evolve  the vacuum condensates and the $\overline{MS}$ mass to the energy scale $\mu$ using the $C\otimes \gamma_\mu C$ type current,   and obtain the mass $4.66\,\rm{GeV}$ or $4.70\,\rm{GeV}$ for the $Y(4660)$.

In Refs.\cite{Wang4660-CTP,Nielsen4660-1110}, the molecule  currents,
\begin{eqnarray}
J_\mu(x)&=&\bar{c}(x)\gamma_\mu c(x)\, \bar{q}(x)q(x)\, ,
\end{eqnarray}
are chosen to study the $Y(4260)$ and $Y(4660)$ in the QCD sum rules, and it is observed that the $Y(4660)$ can be assigned to be the $\psi^{\prime}f_0(980)$ molecular state \cite{Wang4660-CTP}, and the $Y(4260)$ cannot be assigned to be the $J/\psi f_0(980)$ molecular state \cite{Nielsen4660-1110}. Again the parameters are taken as that in  Ref.\cite{WangJPG} and  in Ref.\cite{Nielsen-4260-4460}, respectively.

In Ref.\cite{Nielsen4260-1209}, Dias et al take the $Y(4260)$ as a mixed charmonium-tetraquark state and choose the current $J_\mu(x)$,
 \begin{eqnarray}
J_\mu(x)&=&J^2_\mu(x)\cos\theta +J^4_\mu(x)\sin\theta\, ,
\end{eqnarray}
where
\begin{eqnarray}
J^4_\mu(x)&=&\frac{\epsilon^{ijk}\epsilon^{imn}}{\sqrt{2}}\left\{q^T_j(x)C\gamma_5 c_k(x) \bar{q}_m(x)\gamma_\mu \gamma_5C \bar{c}_n^T(x)+q_j^T(x)C\gamma_5\gamma_\mu c_k(x)\bar{q}_m(x)\gamma_5C \bar{c}_n^T(x) \right\} \, , \nonumber\\
J^2_\mu(x)&=& \frac{1}{\sqrt{2}}\langle\bar{q}q\rangle \,  \bar{c}(x)\gamma_\mu c(x) \, ,
\end{eqnarray}
to study its mass and decay width with the QCD sum rules, and observe that at the mixing angle around $\theta \approx (53.0\pm
0.5)^{\circ}$,   the  mass of the $Y(4260)$ can be reproduced but the  decay width is far below the experimental value.

In this article,  we take the axialvector ($C\gamma_\mu$) diquark states as the basic constituents \cite{WangDiquark,WangLDiquark},  construct the
\begin{eqnarray}
&&C\gamma_\mu \otimes \gamma_\nu C-C\gamma_\nu \otimes \gamma_\mu C\, ,
\end{eqnarray}
type  tensor  current without introducing the additional P-wave  to interpolate both the  vector and axialvector tetraquark states, and study the $Y(4260)$, $Y(4360)$, $Y(4660/4630)$ and $Z_c(4020/4025)$ with the QCD sum rules by calculating the operator product expansion up to the vacuum condensates of dimension 10. The tensor current is expected to couple to the vector tetraquark state with smaller mass compared to the $C\gamma_\alpha \otimes \partial_\mu \gamma^\alpha C$, $C\gamma_5\otimes\partial_\mu \gamma_5C$, $C\otimes \gamma_\mu C$,
$C\gamma_5 \otimes \gamma_5\gamma_\mu C$ type axialvector currents, so as to reproduce the mass of the $Y(4260)$ as the vector tetraquark state. Furthermore, we study the $Z_c^0(4020/4025)$ as the axialvector tetraquark state consists of an  axialvector diquark pair, which is expected to have slight larger mass than the
 $C\gamma_5 \otimes \gamma_\mu C$ type tetraquark state \cite{WangDiquark,WangLDiquark}.  In Ref.\cite{WangHuang-PRD}, we choose the $C\gamma_5 \otimes \gamma_\mu C$ type current to study the axialvector tetraquark states, and obtain the mass $M_{Z_c(3900)}=3.91^{+0.11}_{-0.09}\,\rm{GeV}$ for the  $Z_c(3900)$ with the assignment
 $J^{PC}=1^{+-}$.

The article is arranged as follows:  we derive the QCD sum rules for the masses and pole residues of  the  $Y(4260)$, $Y(4360)$, $Y(4660)$ and $Z_c(4020)$  as pure tetraquark states in section 2; in section 3, we derive the QCD sum rules for the masses and pole residues of  the  $Y(4260)$ and $Y(4360)$ as mixed charmonium-tetraquark states; section 4 is reserved for our conclusion.

\section{QCD sum rules for  the  $Y(4260)$, $Y(4360)$, $Y(4660)$ and $Z_c(4020)$ as pure tetraquark states}
In the following, we write down  the two-point correlation function $\Pi_{\mu\nu\alpha\beta}(p)$  in the QCD sum rules,
\begin{eqnarray}
\Pi_{\mu\nu\alpha\beta}(p)&=&i\int d^4x e^{ip \cdot x} \langle0|T\left\{\eta_{\mu\nu}(x)\eta_{\alpha\beta}^{\dagger}(0)\right\}|0\rangle \, ,
\end{eqnarray}

\begin{eqnarray}
 \eta_{\mu\nu}(x)&=&\frac{\epsilon^{ijk}\epsilon^{imn}}{2}\left\{u^T_j(x)C\gamma_\mu c_k(x) \bar{u}_m(x)\gamma_\nu C \bar{c}^T_n(x)+d^T_j(x)C\gamma_\mu c_k(x) \bar{d}_m(x)\gamma_\nu C \bar{c}^T_n(x)\right.\nonumber\\
 &&\left.-u^T_j(x)C\gamma_\nu c_k(x)\bar{u}_m(x)\gamma_\mu C \bar{c}^T_n(x) -d^T_j(x)C\gamma_\nu c_k(x)\bar{d}_m(x)\gamma_\mu C \bar{c}^T_n(x)\right\} \, ,
\end{eqnarray}
where the $i$, $j$, $k$, $m$, $n$ are color indexes, the $C$ is the charge conjugation matrix.
 The charged partner $\widetilde{\eta}_{\mu\nu}(x)$,
\begin{eqnarray}
 \widetilde{\eta}_{\mu\nu}(x)&=&\frac{\epsilon^{ijk}\epsilon^{imn}}{\sqrt{2}}\left\{u^T_j(x) C\gamma_\mu c_k(x) \bar{d}_m(x) \gamma_\nu C \bar{c}^T_n(x)  -u^T_j(x) C\gamma_\nu c_k(x) \bar{d}_m(x) \gamma_\mu C \bar{c}^T_n(x) \right\} \, ,
 \end{eqnarray}
couples to the $Z_c^+(4020/4025)$ potentially. In the isospin limit, the currents $\eta_{\mu\nu}(x)$ and $\widetilde{\eta}_{\mu\nu}(x)$  couple to the tetraquark states with degenerate masses.

At the hadronic side, we can insert  a complete set of intermediate hadronic states with
the same quantum numbers as the current operator $\eta_{\mu\nu}(x)$  into the
correlation function $\Pi_{\mu\nu\alpha\beta}(p)$ to obtain the hadronic representation
\cite{SVZ79,Reinders85}. After isolating the ground state
contributions of the  axialvector and vector tetraquark states, we get the following results,
\begin{eqnarray}
\Pi_{\mu\nu\alpha\beta}(p)&=&\frac{\lambda_{ Z}^2}{M_{Z}^2-p^2}\left(p^2g_{\mu\alpha}g_{\nu\beta} -p^2g_{\mu\beta}g_{\nu\alpha} -g_{\mu\alpha}p_{\nu}p_{\beta}-g_{\nu\beta}p_{\mu}p_{\alpha}+g_{\mu\beta}p_{\nu}p_{\alpha}+g_{\nu\alpha}p_{\mu}p_{\beta}\right) \nonumber\\
&&+\frac{\lambda_{ Y}^2}{M_{Y}^2-p^2}\left( -g_{\mu\alpha}p_{\nu}p_{\beta}-g_{\nu\beta}p_{\mu}p_{\alpha}+g_{\mu\beta}p_{\nu}p_{\alpha}+g_{\nu\alpha}p_{\mu}p_{\beta}\right) +\cdots \, \, ,
\end{eqnarray}
where the $Z$ denotes the axialvector tetraquark state $Z_c(4020)$, the $Y$ denotes the vector tetraquark state $Y(4260)$, $Y(4360)$ or $Y(4660)$, the  pole residues  $\lambda_{Z}$ and $\lambda_{Y}$ are defined by
\begin{eqnarray}
  \langle 0|\eta_{\mu\nu}(0)|Z_c(p)\rangle &=& \lambda_{Z} \, \epsilon_{\mu\nu\alpha\beta} \, \varepsilon^{\alpha}p^{\beta}\, , \nonumber\\
 \langle 0|\eta_{\mu\nu}(0)|Y(p)\rangle &=& \lambda_{Y} \left(\varepsilon_{\mu}p_{\nu}-\varepsilon_{\nu}p_{\mu} \right)\, ,
\end{eqnarray}
the  $\varepsilon_\mu$ are the polarization vectors of the vector and axialvector tetraquark states  with the following property,
 \begin{eqnarray}
\sum_{\lambda}\varepsilon^*_{\mu}(\lambda,p)\varepsilon_{\nu}(\lambda,p)&=&-g_{\mu\nu}+\frac{p_\mu p_\nu}{p^2} \, .
 \end{eqnarray}
We can rewrite the correlation function $\Pi_{\mu\nu\alpha\beta}(p)$ into the following form according to Lorentz covariance,
\begin{eqnarray}
\Pi_{\mu\nu\alpha\beta}(p)&=&\Pi_Z(p^2)\left(p^2g_{\mu\alpha}g_{\nu\beta} -p^2g_{\mu\beta}g_{\nu\alpha} -g_{\mu\alpha}p_{\nu}p_{\beta}-g_{\nu\beta}p_{\mu}p_{\alpha}+g_{\mu\beta}p_{\nu}p_{\alpha}+g_{\nu\alpha}p_{\mu}p_{\beta}\right) \nonumber\\
&&+\Pi_Y(p^2)\left( -g_{\mu\alpha}p_{\nu}p_{\beta}-g_{\nu\beta}p_{\mu}p_{\alpha}+g_{\mu\beta}p_{\nu}p_{\alpha}+g_{\nu\alpha}p_{\mu}p_{\beta}\right) \, .
\end{eqnarray}

Now we project out the components $\Pi_Z(p^2)$ and $\Pi_Y(p^2)$ by introducing the operators $P_Z^{\mu\nu\alpha\beta}$ and $P_Y^{\mu\nu\alpha\beta}$,
\begin{eqnarray}
\widetilde{\Pi}_Z(p^2)&=&p^2\Pi_Z(p^2)=P_Z^{\mu\nu\alpha\beta}\Pi_{\mu\nu\alpha\beta}(p) \, , \nonumber\\
\widetilde{\Pi}_Y(p^2)&=&p^2\Pi_Y(p^2)=P_Y^{\mu\nu\alpha\beta}\Pi_{\mu\nu\alpha\beta}(p) \, ,
\end{eqnarray}
where
\begin{eqnarray}
P_Z^{\mu\nu\alpha\beta}&=&\frac{1}{6}\left( g^{\mu\alpha}-\frac{p^\mu p^\alpha}{p^2}\right)\left( g^{\nu\beta}-\frac{p^\nu p^\beta}{p^2}\right)\, , \nonumber\\
P_Y^{\mu\nu\alpha\beta}&=&\frac{1}{6}\left( g^{\mu\alpha}-\frac{p^\mu p^\alpha}{p^2}\right)\left( g^{\nu\beta}-\frac{p^\nu p^\beta}{p^2}\right)-\frac{1}{6}g^{\mu\alpha}g^{\nu\beta}\, .
\end{eqnarray}

 In the following,  we carry out  the operator product expansion for the correlation function $\Pi_{\mu\nu\alpha\beta}(p)$ up to the vacuum condensates of dimension 10, and project out the components
  \begin{eqnarray}
\widetilde{\Pi}_Z(p^2)&=&P_Z^{\mu\nu\alpha\beta}\Pi_{\mu\nu\alpha\beta}(p) \, , \nonumber\\
\widetilde{\Pi}_Y(p^2)&=&P_Y^{\mu\nu\alpha\beta}\Pi_{\mu\nu\alpha\beta}(p) \, ,
\end{eqnarray}
at the QCD side, and obtain the QCD spectral densities through dispersion relation,
\begin{eqnarray}
\rho_Z(s)&=&\frac{{\rm Im}\widetilde{\Pi}_Z(s)}{\pi}\, , \nonumber\\
\rho_Y(s)&=&\frac{{\rm Im}\widetilde{\Pi}_Y(s)}{\pi}\, ,
\end{eqnarray}
where we take into account the contributions of the terms $D_0$, $D_3$, $D_4$, $D_5$, $D_6$, $D_7$, $D_8$ and $D_{10}$,
\begin{eqnarray}
D_0&=& {\rm perturbative \,\,\,\, terms}\, , \nonumber\\
D_3&\propto& \langle \bar{q}q\rangle\, ,  \nonumber\\
D_4&\propto&  \langle \frac{\alpha_s GG}{\pi}\rangle\, ,  \nonumber\\
D_5&\propto& \langle \bar{q}g_s\sigma Gq\rangle \, , \nonumber\\
D_6&\propto& \langle \bar{q}q\rangle^2\, , \, g^2_s\langle \bar{q}q\rangle^2  \, , \nonumber\\
D_7&\propto& \langle \bar{q}q\rangle \langle \frac{\alpha_s GG}{\pi}\rangle  \, , \nonumber\\
D_8&\propto& \langle\bar{q}q\rangle\langle \bar{q}g_s\sigma Gq\rangle\, ,  \nonumber\\
D_{10}&\propto& \langle \bar{q}g_s\sigma Gq\rangle^2\, , \, \langle \bar{q}q\rangle^2 \langle \frac{\alpha_s GG}{\pi}\rangle  \, .
\end{eqnarray}
The explicit expressions of the QCD spectral densities $\rho_Z(s)$ and $\rho_Y(s)$  are given in the Appendix.
The four-quark condensate $g_s^2\langle \bar{q}q\rangle^2$ comes from the terms
$\langle \bar{q}\gamma_\mu t^a q g_s D_\eta G^a_{\lambda\tau}\rangle$, $\langle\bar{q}_jD^{\dagger}_{\mu}D^{\dagger}_{\nu}D^{\dagger}_{\alpha}q_i\rangle$  and
$\langle\bar{q}_jD_{\mu}D_{\nu}D_{\alpha}q_i\rangle$, rather than comes from the perturbative corrections of $\langle \bar{q}q\rangle^2$ (see Ref.\cite{WangHuang-PRD} for the technical details).
 The condensates $\langle g_s^3 GGG\rangle$, $\langle \frac{\alpha_s GG}{\pi}\rangle^2$,
 $\langle \frac{\alpha_s GG}{\pi}\rangle\langle \bar{q} g_s \sigma Gq\rangle$ have the dimensions 6, 8, 9 respectively,  but they are   the vacuum expectations
of the operators of the order    $\mathcal{O}( \alpha_s^{3/2})$, $\mathcal{O}(\alpha_s^2)$, $\mathcal{O}( \alpha_s^{3/2})$ respectively, and neglected.  We take
the truncations $n\leq 10$ and $k\leq 1$ in a consistent way,
the operators of the orders $\mathcal{O}( \alpha_s^{k})$ with $k> 1$ are  discarded.  In Table 1 and Table 2, we show the contributions of the vacuum condensates of dimension 4 and 10 explicitly, $|D_4|=1\%$, $(1-2)\%$, $(2-3)\%$, $2\%$ in the Borel windows  for the $Z_c(4020)$, $Y(4660)$, $Y(4260)$, $Y(4360)$, respectively; and $D_{10}\,\ll 1\%$, $\ll 1\%$, $1\leq\%$, $< 1\%$ in the  Borel windows for the $Z_c(4020)$, $Y(4660)$, $Y(4260)$, $Y(4360)$, respectively. Although the vacuum condensates are vacuum expectations of the   operators of the order $\mathcal{O}( \alpha_s)$ both in the terms $D_4$ and $D_{10}$, $|D_4|\gg D_{10}$, as there are additional factors  $\frac{1}{T^2}$, $\frac{1}{T^4}$ and $\frac{1}{T^6}$ in the $D_{10}$, which suppress the contributions greatly. The operators in the condensates $\langle g_s^3 GGG\rangle$, $\langle \frac{\alpha_s GG}{\pi}\rangle^2$,
 $\langle \frac{\alpha_s GG}{\pi}\rangle\langle \bar{q} g_s \sigma Gq\rangle$ are suppressed by additional factors $\mathcal{O}( \alpha_s^{1/2})$, $\mathcal{O}(\alpha_s)$, $\mathcal{O}( \alpha_s^{1/2})$ respectively  and additional factor $\frac{1}{T^2}$ compared with the operator in the $D_4$ or $\langle \frac{\alpha_s GG}{\pi}\rangle$, their  contributions are expected to be of the same order as the $D_{10}$ and neglectable. In Ref.\cite{ZhangJR3900}, Zhang calculates the contributions of the $\langle g_s^3 GGG\rangle$, $\langle \frac{\alpha_s GG}{\pi}\rangle^2$,
 $\langle \frac{\alpha_s GG}{\pi}\rangle\langle \bar{q} g_s \sigma Gq\rangle$ explicitly in the QCD sum rules for the $Z_c(3900)$ as a $\bar{D}D^{*}
$ molecular state, their  contributions are tiny in the Borel window.

 Once the analytical expressions  of the QCD spectral densities   $\rho_Z(s)$ and $\rho_Y(s)$
are obtained,  we can take the
quark-hadron duality below the continuum thresholds  $s_0$ and perform Borel transform  with respect to
the variable $P^2=-p^2$ to obtain  the QCD sum rules:
\begin{eqnarray}
\lambda^2_{Z}M^2_{Z}\, \exp\left(-\frac{M^2_Z}{T^2}\right)= \int_{4m_c^2}^{s_0} ds\, \rho_Z(s) \, \exp\left(-\frac{s}{T^2}\right) \, , \\
\lambda^2_{Y}M^2_{Y}\, \exp\left(-\frac{M^2_Y}{T^2}\right)= \int_{4m_c^2}^{s_0} ds\, \rho_Y(s) \, \exp\left(-\frac{s}{T^2}\right) \, .
\end{eqnarray}

We differentiate   Eqs.(20-21) with respect to  $\frac{1}{T^2}$, eliminate the
 pole residues $\lambda_{Z}$ and $\lambda_{Y}$, and obtain the QCD sum rules for
 the masses of the axialvector and vector  tetraquark states,
 \begin{eqnarray}
 M^2_{Z}&=& \frac{\int_{4m_c^2}^{s_0} ds\,\frac{d}{d \left(-1/T^2\right)}\,\rho_Z(s)\,\exp\left(-\frac{s}{T^2}\right)}{\int_{4m_c^2}^{s_0} ds\, \rho_Z(s)\,\exp\left(-\frac{s}{T^2}\right)}\, , \\
 M^2_{Y}&=& \frac{\int_{4m_c^2}^{s_0} ds\,\frac{d}{d \left(-1/T^2\right)}\,\rho_Y(s)\,\exp\left(-\frac{s}{T^2}\right)}{\int_{4m_c^2}^{s_0} ds\, \rho_Y(s)\,\exp\left(-\frac{s}{T^2}\right)}\, .
\end{eqnarray}

We take  the standard values of the vacuum condensates,
$\langle\bar{q}q \rangle=-(0.24\pm 0.01\, \rm{GeV})^3$,
$\langle\bar{q}g_s\sigma G q \rangle=m_0^2\langle \bar{q}q \rangle$,
$m_0^2=(0.8 \pm 0.1)\,\rm{GeV}^2$, $\langle \frac{\alpha_s
GG}{\pi}\rangle=(0.33\,\rm{GeV})^4 $    at the energy scale  $\mu=1\, \rm{GeV}$
\cite{SVZ79,Reinders85}.
The quark condensate and mixed quark condensate evolve with the   renormalization group equation,
$\langle\bar{q}q \rangle(\mu)=\langle\bar{q}q \rangle(Q)\left[\frac{\alpha_{s}(Q)}{\alpha_{s}(\mu)}\right]^{\frac{4}{9}}$ and
 $\langle\bar{q}g_s \sigma Gq \rangle(\mu)=\langle\bar{q}g_s \sigma Gq \rangle(Q)\left[\frac{\alpha_{s}(Q)}{\alpha_{s}(\mu)}\right]^{\frac{2}{27}}$.

In the article, we take the $\overline{MS}$ mass $m_{c}(m_c)=(1.275\pm0.025)\,\rm{GeV}$
 from the Particle Data Group \cite{PDG}, and take into account
the energy-scale dependence of  the $\overline{MS}$ mass from the renormalization group equation,
\begin{eqnarray}
m_c(\mu)&=&m_c(m_c)\left[\frac{\alpha_{s}(\mu)}{\alpha_{s}(m_c)}\right]^{\frac{12}{25}} \, ,\nonumber\\
\alpha_s(\mu)&=&\frac{1}{b_0t}\left[1-\frac{b_1}{b_0^2}\frac{\log t}{t} +\frac{b_1^2(\log^2{t}-\log{t}-1)+b_0b_2}{b_0^4t^2}\right]\, ,
\end{eqnarray}
  where $t=\log \frac{\mu^2}{\Lambda^2}$, $b_0=\frac{33-2n_f}{12\pi}$, $b_1=\frac{153-19n_f}{24\pi^2}$, $b_2=\frac{2857-\frac{5033}{9}n_f+\frac{325}{27}n_f^2}{128\pi^3}$,  $\Lambda=213\,\rm{MeV}$, $296\,\rm{MeV}$  and  $339\,\rm{MeV}$ for the flavors  $n_f=5$, $4$ and $3$, respectively  \cite{PDG}.

In previous works, we described   the hidden charm (or bottom) four-quark systems  $q\bar{q}^{\prime}Q\bar{Q}$
by a double-well potential \cite{WangTetraquarkCTP,WangEPJC-4660,Wang-Tetraquark-DW, Wang-molecule,Wang-Octet}.     In the four-quark system $q\bar{q}^{\prime}Q\bar{Q}$,
 the $Q$-quark serves as a static well potential and  combines with the light quark $q$  to form a heavy diquark $\mathcal{D}^i_{qQ}$ in  color antitriplet $q+Q \to  \mathcal{D}^i_{qQ}$ \cite{WangTetraquarkCTP,WangEPJC-4660,Wang-Tetraquark-DW},
or combines with the light antiquark $\bar{q}^\prime$ to form a heavy meson in color singlet (meson-like state in color octet) $\bar{q}^\prime+Q \to  \bar{q}^{\prime}Q\,\, (\bar{q}^{\prime}\lambda^{a}Q)$ \cite{Wang-molecule,Wang-Octet};
 the $\bar{Q}$-quark serves  as another static well potential and combines with the light antiquark $\bar{q}^\prime$  to form a heavy antidiquark $\mathcal{D}^i_{\bar{q}^{\prime}\bar{Q}}$ in  color triplet $\bar{q}^{\prime}+\bar{Q} \to  \mathcal{D}^i_{\bar{q}^{\prime}\bar{Q}}$ \cite{WangTetraquarkCTP,WangEPJC-4660,Wang-Tetraquark-DW},
or combines with the light quark $q$ to form a heavy meson in color singlet (meson-like state in color octet)
$q+\bar{Q} \to  \bar{Q}q\,\, (\bar{Q}\lambda^{a}q)$ \cite{Wang-molecule,Wang-Octet},
where the $i$ is color index, the $\lambda^a$ is Gell-Mann matrix.
 Then
\begin{eqnarray}
 \mathcal{D}^i_{qQ}+\mathcal{D}^i_{\bar{q}^{\prime}\bar{Q}} &\to &  {\rm compact \,\,\, tetraquark \,\,\, states}\, , \nonumber\\
 \bar{q}^{\prime}Q+\bar{Q}q &\to & {\rm loose  \,\,\, molecular \,\,\, states}\, , \nonumber\\
  \bar{q}^{\prime}\lambda^aQ+\bar{Q}\lambda^a q &\to & {\rm   molecule-like  \,\,\, states}\, ,
\end{eqnarray}
the two heavy quarks $Q$ and $\bar{Q}$ stabilize the four-quark systems $q\bar{q}^{\prime}Q\bar{Q}$, just as in the case
of the $(\mu^-e^+)(\mu^+ e^-)$ molecule in QED \cite{Brodsky-2014}.

In Refs.\cite{WangTetraquarkCTP,WangEPJC-4660,WangHuang-PRD,Wang-Tetraquark-DW, Wang-molecule,Wang-Octet}, we study the acceptable energy scales of the QCD spectral densities  for the hidden  charm (bottom) four-quark systems $q\bar{q}^{\prime}Q\bar{Q}$ with  the QCD sum rules in details for the first time,  and suggest a  formula,
\begin{eqnarray}
\mu &=&\sqrt{M^2_{X/Y/Z}-(2{\mathbb{M}}_Q)^2} \, ,
 \end{eqnarray}
to determine  the energy scales, where the $X$, $Y$, $Z$ denote the four-quark systems, and the ${\mathbb{M}}_Q$ denotes the effective heavy quark masses.
In Refs.\cite{WangHuang-PRD,Wang-Tetraquark-DW}, we obtain the optimal value of the effective mass  for  the diquark-antidiquark type tetraquark states,   ${\mathbb{M}}_c=1.8\,\rm{GeV}$. Recently, we re-checked the numerical calculations and found that there exists  a small error involving the mixed condensates.  The Borel windows are modified slightly and the numerical results are also improved slightly after the small error is corrected, the conclusions survive, the optimal value of the effective mass is ${\mathbb{M}}_c=1.82\,\rm{GeV}$  for  the diquark-antidiquark type tetraquark states. In this article, we choose the value ${\mathbb{M}}_c=1.82\,\rm{GeV}$.

Firstly, we assume that  the $Y(4260)$ and $Y(4360)$   are the ground state vector tetraquark states, the energy gap between the ground states and
 the first radial excited states is about $(0.4-0.6)\,\rm{GeV}$, just like that of the conventional mesons.
In case I,  the $Y(4260)$ is the ground state vector tetraquark state;
in case II,  the $Y(4360)$ is the ground state vector tetraquark state.

In Fig.1, we plot the  masses of the vector  tetraquark states    with variations of the  Borel parameters $T^2$ and energy scales $\mu$ for the continuum  threshold parameters $s_{Y(4260)}^0=23\,\rm{GeV}^2$ and $s_{Y(4360)}^0=24\,\rm{GeV}^2$, respectively.
According to the formula in Eq.(26),   the  energy scales $\mu_{Y(4260)}=2.2\,\rm{GeV}$ and $\mu_{Y(4360)}=2.4\,\rm{GeV}$ are the optimal energy scales.
From Fig.1, we can see that the masses decrease monotonously with increase of the energy scales at the value $T^2> 2.7\,\rm{GeV}^2$. However, it is impossible to reproduce the experimental values even if much larger energy scales are taken, the QCD sum rules do not support   assigning the $Y(4260)$ and $Y(4360)$  to be  the  vector tetraquark states.

\begin{figure}
\centering
\includegraphics[totalheight=6cm,width=7cm]{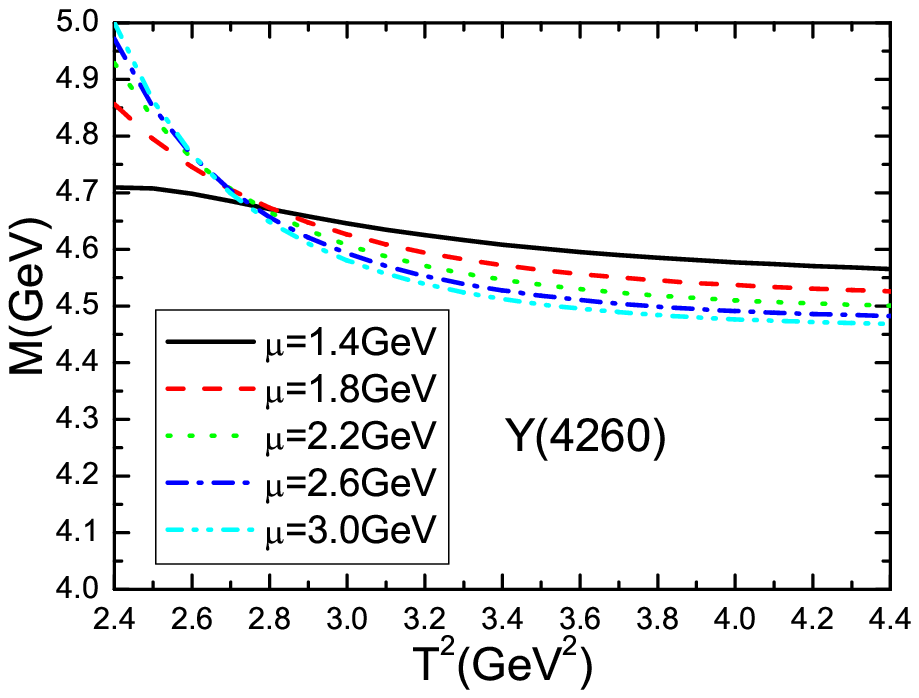}
\includegraphics[totalheight=6cm,width=7cm]{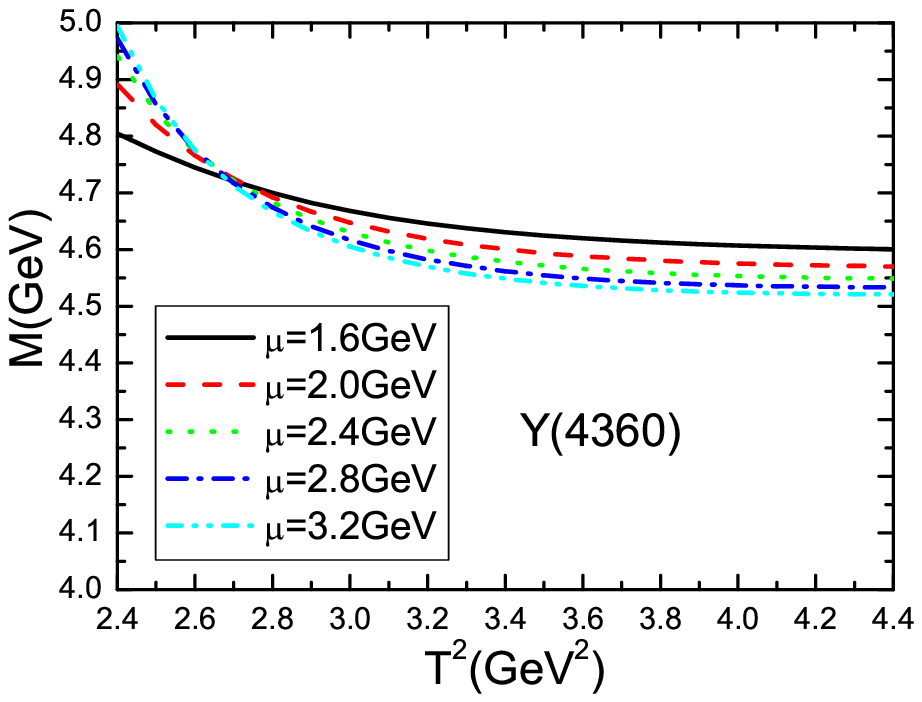}
  \caption{ The predicted masses  with variations of the  Borel parameters $T^2$ and the energy scales $\mu$. }
\end{figure}

In the conventional QCD sum rules \cite{SVZ79,Reinders85}, there are two criteria (pole dominance at the phenomenological side and convergence of the operator product
expansion) for choosing  the Borel parameters $T^2$ and continuum threshold parameters $s_0$.  Now we assume the tensor current couples potentially to the vector tetraquark state $Y(4660)$ and the axialvector tetraquark state $Z_c(4020)$, and search for  the Borel parameters $T^2$ and continuum threshold parameters $s_0$.
 The resulting Borel parameters, continuum threshold parameters, energy scales, pole contributions  and contributions of the vacuum condensates of dimension 10  are shown in Table 1.

\begin{table}
\begin{center}
\begin{tabular}{|c|c|c|c|c|c|c|c|c|}\hline\hline
              & $T^2 (\rm{GeV}^2)$ & $s_0 (\rm{GeV}^2)$ & $\mu(\rm{GeV})$   & pole         & $|D_{4}|$    & $D_{10}$          \\ \hline
$Z_c(4020)$   & $3.2-3.6$          & $21.0\pm1.0$       & $1.7$             & $(40-61)\%$  & $ 1\%$       & $\ll 1\%$         \\ \hline
$Y(4660)$     & $3.5-3.9$          & $26.5\pm1.0$       & $2.9$             & $(46-64)\%$  & $(1-2)\%$    & $\ll 1\%$         \\ \hline
 \hline
\end{tabular}
\end{center}
\caption{ The Borel parameters, continuum threshold parameters, energy scales, pole contributions  and contributions of the vacuum condensates of dimension 4 and 10 for the $Z_c(4020)$ and $Y(4660)$. }
\end{table}

\begin{figure}
\centering
\includegraphics[totalheight=6cm,width=7cm]{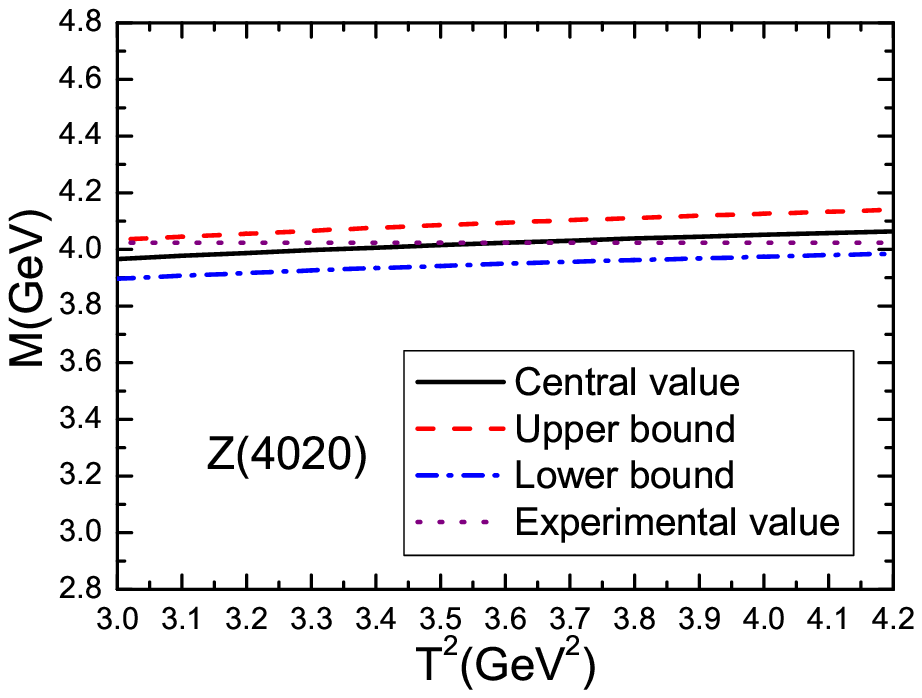}
\includegraphics[totalheight=6cm,width=7cm]{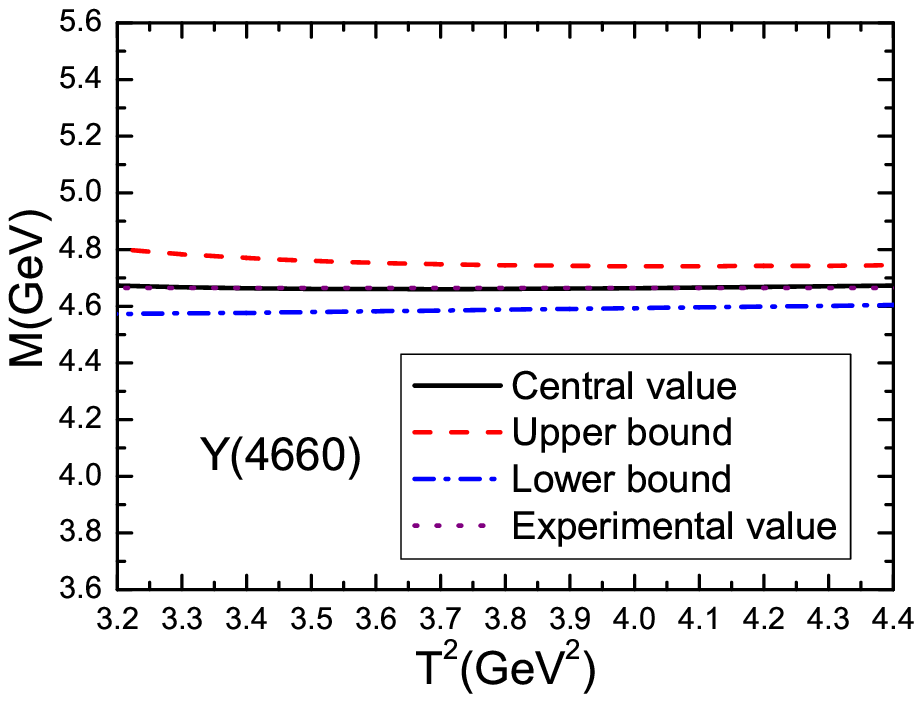}
  \caption{ The masses  with variations of the  Borel parameters $T^2$ for  the tetraquark states $Z_c(4020)$ and $Y(4660)$. }
\end{figure}

Then we take into account all uncertainties of the input parameters,
and obtain the values of the masses (and pole residues) of
 the    axialvector and vector tetraquark states, which are  shown  in Fig.2,
 \begin{eqnarray}
 M_{Z_c(4020)}&=&\left(4.01\pm0.08\right) \, \rm{GeV}\, , \nonumber\\
 \lambda_{Z_c(4020)}&=&\left(7.31\pm0.99\right)\times 10^{-3} \, \rm{GeV}^4\, , \\
 M_{Y(4660)}&=&\left(4.66\pm0.09\right) \, \rm{GeV}\, , \nonumber\\
 \lambda_{Y(4660)}&=&(1.33\pm0.15)\times 10^{-2} \, \rm{GeV}^4\, .
 \end{eqnarray}
The present prediction  $ M_{Z_c(4020)}=\left(4.01\pm0.08\right) \, \rm{GeV}$ is consistent with the experimental values $M_{Z_c^\pm(4025)}=(4026.3\pm2.6\pm3.7)\,\rm{MeV}$, $M_{Z_c^\pm(4020)}=(4022.9\pm 0.8\pm 2.7)\,\rm{MeV}$, $M_{Z_c^0(4020)}=(4023.9\pm2.2\pm3.8)\,\rm{MeV}$, $M_{Z_c^0(4025)}=(4025.5^{+2.0}_{-4.7}\pm3.1)\,\rm{MeV}$
from the BESIII collaboration \cite{BES1308,BES1309,BES1409,BES1507},    which favors assigning the $Z_c(4020/4025)$   to be  the $J^{PC}=1^{+-}$    diquark-antidiquark type tetraquark state. In Ref.\cite{WangTetraquarkCTP}, the contributions of the vector and axialvector tetraquark states are not separated explicitly, the prediction $M_{Z_c(4020/4025)} =\left(4.02^{+0.07}_{-0.08}\right)\,\rm{GeV}$ is consistent with the present value
$ M_{Z_c(4020)}=\left(4.01\pm0.08\right) \, \rm{GeV}$, which indicates  the contamination from the vector tetraquark state $Y(4660)$ is small, as the energy gap $M_{Y(4660)}-M_{Z_c(4020)}\approx 0.65\,\rm{GeV}$.
The present prediction  $M_{Y(4660)}=\left(4.66\pm0.09\right) \, \rm{GeV}$ is consistent with the experimental value $M_{Y(4660)}=(4665\pm10)\,\rm{MeV}$ within uncertainty \cite{PDG}, which favors assigning the $Y(4660)$   to be  the vector  diquark-antidiquark type tetraquark state.

Now we can see that all the three diquark-antidiquark type currents
$C\otimes \gamma_\mu C$, $C\gamma_5 \otimes \gamma_5\gamma_\mu C$ \cite{WangJPG,Nielsen-4260-4460,WangEPJC-4660},
$C\gamma_\mu \otimes \gamma_\nu C-C\gamma_\nu \otimes \gamma_\mu C$,
couple potentially to the vector tetraquark state $Y(4660)$. In Ref.\cite{ChenZhu}, Chen and Zhu observe that the $C\gamma^\nu\otimes \sigma_{\mu\nu}C$ type current also couples  potentially to the $Y(4660)$.
The interpolating currents of the types
\begin{eqnarray}
&&C\otimes  \gamma_\mu C\, , \nonumber\\
&&C\gamma_5 \otimes \gamma_5\gamma_\mu C\, , \nonumber\\
&&C\gamma^\nu\otimes \sigma_{\mu\nu}C\, ,
\end{eqnarray}
have unstable diquarks, such as the pseudoscalar $C$, vector $C\gamma_\mu \gamma_5$, tensor $C\sigma_{\mu\nu}$ diquarks, and couple potentially to the tetraquark states with the additional P-wave \cite{Maiani-4260}.  In this article, we observe that the $C\gamma_\mu \otimes \gamma_\nu C-C\gamma_\nu \otimes \gamma_\mu C$ type current without unstable diquarks also couples potentially to the vector tetraquark state with the additional P-wave, however, the large mass $\left(4.66\pm0.09\right) \, \rm{GeV}$ disfavors assigning the $Y(4260)$ and $Y(4360)$ to be the vector tetraquark states.
In Ref.\cite{Maiani-4260}, the $Y(4260)$ is identified as the $C\gamma_5\otimes \gamma_5 C$ type vector tetraquark state with an additional P-wave.
 On the other hand, we can also construct
the $C\gamma_\alpha \otimes \partial_\mu \gamma^\alpha C$ type and $C\gamma_5\otimes\partial_\mu \gamma_5C$ type diquark-antidiquark  currents to interpolate the vector tetraquark states \cite{ZhangHuang-PRD,ZhangHuang-JHEP}.

\section{QCD sum rules for  the  $Y(4260)$ and $Y(4360)$ as mixed charmonium-tetraquark states}

Now we take the $Y(4260)$ and $Y(4360)$ to be the mixed charmonium-tetraquark states, and study the masses and pole residues with the QCD sum rules. Firstly, let us write down the interpolating current,
\begin{eqnarray}
J_{\mu\nu}(x)&=&\eta_{\mu\nu}(x)\,\cos\theta +\frac{i}{3}\,\langle\bar{q}q\rangle\, \bar{c}(x)\sigma_{\mu\nu}c(x)\,\sin\theta \, ,
\end{eqnarray}
where the $\theta$ is the mixing angle, the   $\frac{i}{3}\langle\bar{q}q\rangle$ is normalization factor \cite{MixingFactor}. The calculations can be carried out  straightforwardly  with the simple replacement,
\begin{eqnarray}
\eta_{\mu\nu}(x) &\to & J_{\mu\nu}(x) \, ,
\end{eqnarray}
in the correlation function $\Pi_{\mu\nu\alpha\beta}(p)$ in Eq.(1).
The resulting QCD sum rules are
\begin{eqnarray}
\lambda^2_{Y}M^2_{Y}\, \exp\left(-\frac{M^2_Y}{T^2}\right)= \int_{4m_c^2}^{s_0} ds\,\left[ \cos^2\theta \,\rho_Y(s)+2\sin\theta \cos\theta\,\rho_{m}(s)+\sin^2\theta \,\rho_{2}(s) \right] \, \exp\left(-\frac{s}{T^2}\right) \, ,
\end{eqnarray}
where the $\rho_Y(s)$ is the QCD spectral density of the tetraquark component shown in Eq.(18), and
\begin{eqnarray}
\rho_m(s)&=&\rho_2(s)+\frac{\langle\bar{q}q\rangle\langle\bar{q}g_s\sigma Gq\rangle }{144\pi^2}\int_{y_i}^{y_f}dy \left[1+\frac{\widetilde{m}_c^2}{2}\delta(s-\widetilde{m}_c^2) \right]  \, ,
\end{eqnarray}

\begin{eqnarray}
\rho_2(s)&=&\frac{\langle\bar{q}q\rangle^2}{12\pi^2}\int_{y_i}^{y_f}dy \left[y(1-y)s+m_c^2 \right] \nonumber\\
&&+\frac{\langle\bar{q}q\rangle^2}{72} \langle\frac{\alpha_sGG}{\pi}\rangle\int_{0}^{1}dy \left(\frac{1}{3}-\frac{\widetilde{m}_c^2}{T^2}\right)\delta(s-\widetilde{m}_c^2) \nonumber\\
&&+\frac{m_c^2\langle\bar{q}q\rangle^2}{72T^2} \langle\frac{\alpha_sGG}{\pi}\rangle\int_{0}^{1}dy \left[\frac{1}{y^2}+\frac{1}{(1-y)^2}\right]\delta(s-\widetilde{m}_c^2) \nonumber\\
&&+\frac{m_c^2\langle\bar{q}q\rangle^2}{108T^2} \langle\frac{\alpha_sGG}{\pi}\rangle\int_{0}^{1}dy \left[\frac{1-y}{y^2}+\frac{y}{(1-y)^2}\right]\left(1-\frac{\widetilde{m}_c^2}{T^2}\right)\delta(s-\widetilde{m}_c^2) \, ,
\end{eqnarray}
$y_{f}=\frac{1+\sqrt{1-4m_c^2/s}}{2}$, $y_{i}=\frac{1-\sqrt{1-4m_c^2/s}}{2}$,
$ \widetilde{m}_c^2=\frac{m_c^2}{y(1-y)}$, $\int_{y_i}^{y_f}dy \to \int_{0}^{1}dy$,  when the $\delta$ function $\delta\left(s-\widetilde{m}_c^2\right)$ appears.

We differentiate   Eq.(32) with respect to  $\frac{1}{T^2}$, then eliminate the
 pole residues $\lambda_{Y}$, and obtain the QCD sum rules for
 the masses of the mixed charmonium-tetraquark states,
 \begin{eqnarray}
 M^2_{Y}&=& \frac{\int_{4m_c^2}^{s_0} ds\frac{d}{d \left(-1/T^2\right)}\left[ \cos^2\theta \,\rho_Y(s)+2\sin\theta \cos\theta\,\rho_{m}(s)+\sin^2\theta\, \rho_{2}(s) \right]\exp\left(-\frac{s}{T^2}\right)}{\int_{4m_c^2}^{s_0} ds \left[ \cos^2\theta \,\rho_Y(s)+2\sin\theta \cos\theta\,\rho_{m}(s)+\sin^2\theta\, \rho_{2}(s) \right]\exp\left(-\frac{s}{T^2}\right)}\, .
\end{eqnarray}

In case I, we take the $Y(4260)$ as the ground state  mixed charmonium-tetraquark state, and choose the optimal energy scale $\mu=2.2\,\rm{GeV}$.
In case II, we take the $Y(4360)$ as the ground state  mixed charmonium-tetraquark state, and choose the optimal energy scale $\mu=2.4\,\rm{GeV}$.
Then we impose the  two criteria (pole dominance at the phenomenological side and convergence of the operator product
expansion) of the QCD sum rules on the $Y(4260)$ and $Y(4360)$, and  search for   the mixing angles $\theta$, Borel parameters $T^2$ and continuum threshold parameters  $s_0$.  The resulting  mixing angles, Borel parameters, continuum threshold parameters, energy scales, pole contributions  and contributions of the vacuum condensates of dimension 10  are shown in Table 2. From the table, we can see that the  two criteria of the conventional QCD sum rules can be satisfied, so we expect to make reasonable predictions.

\begin{table}
\begin{center}
\begin{tabular}{|c|c|c|c|c|c|c|c|}\hline\hline
            & $\theta$      & $T^2 (\rm{GeV}^2)$ & $s_0 (\rm{GeV}^2)$  & $\mu(\rm{GeV})$ & pole         & $|D_{4}|$     & $D_{10}$          \\ \hline
$Y(4260)$   & $5.84^\circ$  & $2.9-3.3$          & $23.0\pm1.0$        & $2.2$           & $(40-63)\%$  & $(2-3)\%$     & $\leq 1\%$        \\ \hline
$Y(4360)$   & $5.61^\circ$  & $3.1-3.5$          & $24.0\pm1.0$        & $2.4$           & $(42-64)\%$  & $2\%$         & $< 1\%$           \\ \hline
 \hline
\end{tabular}
\end{center}
\caption{ The mixing angles, Borel parameters, continuum threshold parameters, energy scales, pole contributions  and contributions of the vacuum condensates of dimension 4 and 10 for the $Y(4260)$ and $Y(4360)$.}
\end{table}

\begin{figure}
\centering
\includegraphics[totalheight=6cm,width=7cm]{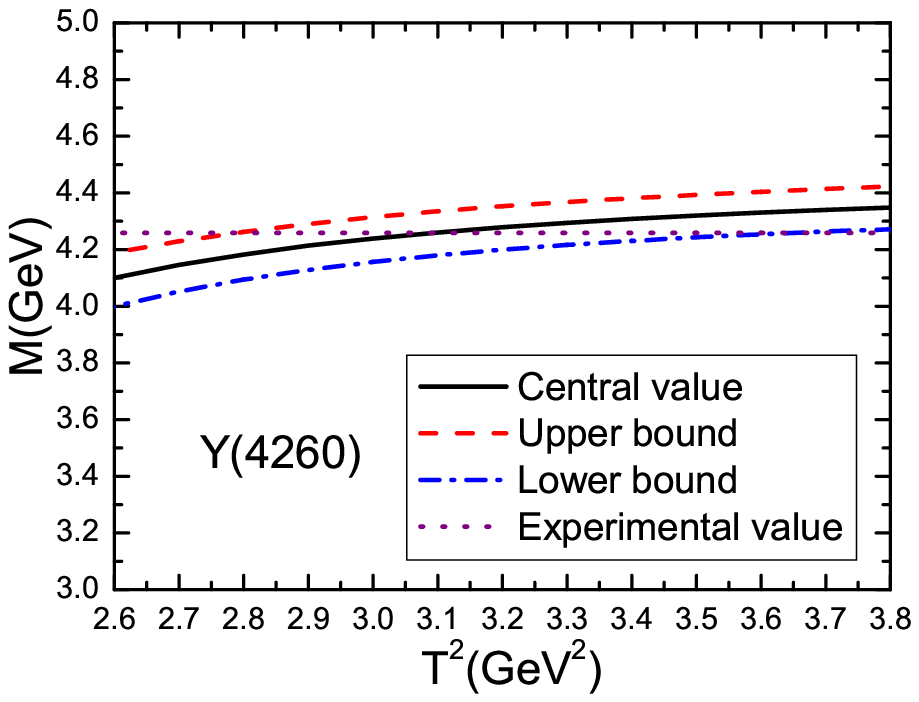}
\includegraphics[totalheight=6cm,width=7cm]{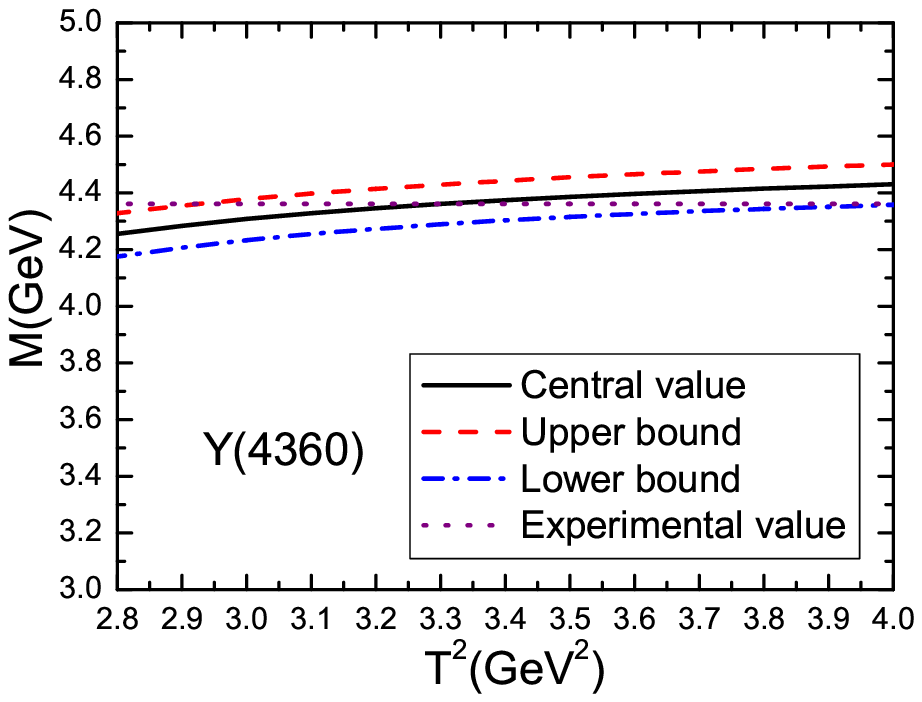}
  \caption{ The masses  with variations of the  Borel parameters $T^2$ for the $Y(4260)$ and $Y(4360)$ as mixed charmonium-tetraquark states. }
\end{figure}

We take into account all uncertainties of the input parameters,
and obtain the values of the masses (and pole residues) of
 the   $Y(4260)$ and $Y(4360)$ as mixed charmonium-tetraquark states, which are  shown explicitly in Fig.3,
 \begin{eqnarray}
  M_{Y(4260)}&=&\left(4.26\pm0.11\right) \, \rm{GeV}\, , \nonumber\\
 \lambda_{Y(4260)}&=&(6.72\pm1.33)\times 10^{-3} \, \rm{GeV}^4\, , \\
 M_{Y(4360)}&=&\left(4.36\pm0.10\right) \, \rm{GeV}\, , \nonumber\\
 \lambda_{Y(4360)}&=&(8.32\pm1.36)\times 10^{-3} \, \rm{GeV}^4\, .
 \end{eqnarray}
The prediction  $ M_{Y(4260)}=\left(4.26\pm0.11\right) \, \rm{GeV}$ is consistent with the experimental value $M_{Y(4260)}=\left(4259 \pm 8 {}^{+2}_{-6}\right) \,\rm{MeV}$ \cite{BaBar4260-0506},   which favors assigning the $Y(4260)$  to be  the mixed charmonium-tetraquark state.
On the other hand, the prediction  $M_{Y(4360)}=\left(4.36\pm0.10\right) \, \rm{GeV}$ is consistent with the experimental value  $M_{Y(4360)}=(4361\pm 9\pm 9)\, \rm{MeV}$ \cite{Belle4660-0707}, which also favors assigning the $Y(4360)$ to be the mixed charmonium-tetraquark state. In the two cases, $\cos^2\theta\approx0.99$, the dominant components  are the tetraquark states, $2\sin\theta \cos\theta\approx 0.20\,\, {\rm or} \,\,0.19$, the mixing effects are also considerable. In Ref.\cite{Nielsen-4260-4460}, the tetraquark component of the $Y(4260)$ is about $\sin^2\theta\approx 0.64$, the conclusion is quite different from the present work. The difference maybe originate from the interpolating  currents and the truncation of the operator product expansion.

\section{Conclusion}
In this article, we construct the axialvector-diquark-axialvector-antidiquark type tensor current to interpolate both the vector and axialvector tetraquark states,  then calculate the contributions of the vacuum condensates up to dimension-10  in the operator product expansion, and obtain the QCD sum rules for  both the vector and axialvector tetraquark states. In calculations,  we use the  formula $\mu=\sqrt{M^2_{X/Y/Z}-(2{\mathbb{M}}_c)^2}$ suggested in our previous work to determine  the energy scales of the QCD spectral densities,  which works well. The numerical results support assigning the $Z_c(4020/4025)$ to be the $J^{PC}=1^{+-}$ diquark-antidiquark type tetraquark state, and  assigning the $Y(4660)$ to be the $J^{PC}=1^{--}$ diquark-antidiquark type tetraquark state. Furthermore, we take the $Y(4260)$ and $Y(4360)$ as the mixed charmonium-tetraquark states, introduce the mixing angle and construct the two-quark-tetraquark type tensor currents to study the masses and pole residues. The experimental values of the masses can be reproduced with suitable mixing angles, the QCD sum rules support assigning the $Y(4260)$ and $Y(4360)$ to be the mixed charmonium-tetraquark states.

\section*{Appendix}

The QCD spectral densities $\rho_Y(s)$ and  $\rho_Z(s)$,
\begin{eqnarray}
\rho_{Y}(s)&=&\frac{1}{6144\pi^6}\int dydz \, yz(1-y-z)^3\left(s-\overline{m}_c^2\right)^2\left(35s^2-18s\overline{m}_c^2-\overline{m}_c^4 \right)  \nonumber\\
&&+\frac{1}{6144\pi^6}\int dydz \, yz(1-y-z)^2\left(s-\overline{m}_c^2\right)^3\left(9s-\overline{m}_c^2\right)  \nonumber\\
&&+\frac{m_c\langle \bar{q}q\rangle}{32\pi^4}\int dydz \, (y+z)(1-y-z)\left(s-\overline{m}_c^2\right)^2  \nonumber \\
&&-\frac{m_c^2}{4608\pi^4} \langle\frac{\alpha_s GG}{\pi}\rangle\int dydz \left( \frac{z}{y^2}+\frac{y}{z^2}\right)(1-y-z)^3 \left\{ 4s+\overline{m}_c^2+\frac{4}{3}\,s^2\,\delta\left(s-\overline{m}_c^2\right)\right\} \nonumber\\
&&-\frac{m_c^2}{4608\pi^4} \langle\frac{\alpha_s GG}{\pi}\rangle\int dydz \left( \frac{z}{y^2}+\frac{y}{z^2}\right)(1-y-z)^2 \left(3s-\overline{m}_c^2\right) \nonumber\\
&&+\frac{1}{18432\pi^4}\langle\frac{\alpha_s GG}{\pi}\rangle\int dydz \, (y+z)   (1-y-z)^2 \left(95s^2-120s\overline{m}_c^2+33\overline{m}_c^4 \right) \nonumber\\
&&+\frac{1}{9216\pi^4}\langle\frac{\alpha_s GG}{\pi}\rangle\int dydz \, (y+z)   (1-y-z)  \left(s-\overline{m}_c^2 \right)\left(5s-\overline{m}_c^2 \right) \nonumber\\
&&-\frac{1}{4608\pi^4}\langle\frac{\alpha_s GG}{\pi}\rangle\int dydz \, (y+z)   (1-y-z)^2  \left(s-\overline{m}_c^2 \right)\left(5s-3\overline{m}_c^2 \right) \nonumber\\
&&+\frac{1}{82944\pi^4}\langle\frac{\alpha_s GG}{\pi}\rangle\int dydz \,    (1-y-z)^3 \left(35s^2-24s\overline{m}_c^2-3\overline{m}_c^4 \right) \nonumber\\
&&-\frac{1}{4608\pi^4}\langle\frac{\alpha_s GG}{\pi}\rangle\int dydz \, yz   (1-y-z)  \left(s-\overline{m}_c^2 \right)\left(5s-3\overline{m}_c^2 \right) \nonumber\\
&&-\frac{1}{27648\pi^4}\langle\frac{\alpha_s GG}{\pi}\rangle\int dydz \,     (1-y-z)^2  \left(s-\overline{m}_c^2 \right)\left(13s-5\overline{m}_c^2 \right) \nonumber\\
&&-\frac{1}{4608\pi^4}\langle\frac{\alpha_s GG}{\pi}\rangle\int dydz \, yz     \left(s-\overline{m}_c^2 \right)^2 \nonumber\\
&&-\frac{m_c\langle \bar{q}g_s\sigma Gq\rangle}{64\pi^4}\int dydz  \, (y+z) \left(s-\overline{m}_c^2 \right) \nonumber\\
&&+\frac{m_c\langle \bar{q}g_s\sigma Gq\rangle}{288\pi^4}\int dydz  \, (1-y-z) \left(s-\overline{m}_c^2 \right) \nonumber\\
&&-\frac{m_c^2\langle\bar{q}q\rangle^2}{12\pi^2}\int dy \nonumber\\
&&+\frac{g_s^2\langle\bar{q}q\rangle^2}{1296\pi^4}\int dydz\,yz\left\{ \left(4+\overline{m}_c^2 \right)+ \frac{4}{3}\,s^2\,\delta\left(s-\overline{m}_c^2 \right)\right\} \nonumber\\
&&+\frac{g_s^2\langle\bar{q}q\rangle^2}{3888\pi^4}\int dy\,y(1-y)\left(3s-\overline{m}_c^2 \right)\nonumber\\
&&+\frac{g_s^2\langle\bar{q}q\rangle^2}{3888\pi^4}\int dydz\,(1-y-z)\left\{ \left(\frac{z}{y}+\frac{y}{z} \right)9\left(s-\overline{m}_c^2 \right)+\left(\frac{z}{y^2}+\frac{y}{z^2} \right)\right.\nonumber\\
&&\left.m_c^2\left[ 5+s\,\delta\left(s-\overline{m}_c^2 \right)\right]-(y+z)\left[\left(24s-6\overline{m}_c^2 \right)+4s^2\,\delta\left(s-\overline{m}_c^2 \right)\right]\right\} \nonumber\\
&&-\frac{g_s^2\langle\bar{q}q\rangle^2}{11664\pi^4}\int dydz\,(1-y-z)\left\{  \left(\frac{z}{y}+\frac{y}{z} \right)9\left(3s-\overline{m}_c^2 \right)+\left(\frac{z}{y^2}+\frac{y}{z^2} \right)\right. \nonumber\\
&&\left.m_c^2\left[ 4+8s\,\delta\left(s-\overline{m}_c^2\right)\right]+(y+z)\left[12s+3\overline{m}_c^2 +4s^2\,\delta\left(s-\overline{m}_c^2\right)\right]\right\} \nonumber
\end{eqnarray}
\begin{eqnarray}
&&-\frac{m_c^3\langle\bar{q}q\rangle}{288\pi^2}\langle\frac{\alpha_sGG}{\pi}\rangle\int dydz \left(\frac{y}{z^3}+\frac{z}{y^3}+\frac{1}{y^2}+\frac{1}{z^2}\right)(1-y-z)\,\delta\left(s-\overline{m}_c^2\right)\nonumber\\
&&+\frac{m_c\langle\bar{q}q\rangle}{96\pi^2}\langle\frac{\alpha_sGG}{\pi}\rangle\int dydz \left(\frac{y}{z^2}+\frac{z}{y^2}\right)(1-y-z) \nonumber\\
&&-\frac{m_c\langle\bar{q}q\rangle}{96\pi^2}\langle\frac{\alpha_sGG}{\pi}\rangle\int dydz \left\{1+\frac{2}{3}\,s\,\delta\left(s-\overline{m}_c^2 \right)\right\} \nonumber\\
&&+\frac{m_c\langle\bar{q}q\rangle}{864\pi^2}\langle\frac{\alpha_sGG}{\pi}\rangle\int dydz \left\{\frac{y}{z}+\frac{z}{y}+\left(\frac{1}{y}+\frac{1}{z}\right)(1-y-z)\right\} \nonumber\\
&&+\frac{m_c\langle\bar{q}q\rangle}{576\pi^2}\langle\frac{\alpha_sGG}{\pi}\rangle\int dy \nonumber\\
&&\frac{m_c^2\langle\bar{q}q\rangle\langle\bar{q}g_s\sigma Gq\rangle}{24\pi^2}\int_0^1 dy \left(1+\frac{s}{T^2} \right)\delta\left(s-\widetilde{m}_c^2\right)\nonumber\\
&&-\frac{\langle\bar{q}q\rangle\langle\bar{q}g_s\sigma Gq\rangle}{432\pi^2}\int_0^1 dy \,s\,\delta\left(s-\widetilde{m}_c^2\right) \nonumber\\
&&-\frac{m_c^2\langle\bar{q}g_s\sigma Gq\rangle^2}{192\pi^2T^6}\int_0^1 dy \, s^2 \, \delta \left( s-\widetilde{m}_c^2\right)
\nonumber\\
&&+\frac{m_c^4\langle\bar{q}q\rangle^2}{216T^4}\langle\frac{\alpha_sGG}{\pi}\rangle\int_0^1 dy  \left\{ \frac{1}{y^3}+\frac{1}{(1-y)^3}\right\} \delta\left( s-\widetilde{m}_c^2\right)\nonumber\\
&&-\frac{m_c^2\langle\bar{q}q\rangle^2}{72T^2}\langle\frac{\alpha_sGG}{\pi}\rangle\int_0^1 dy  \left\{ \frac{1}{y^2}+\frac{1}{(1-y)^2}\right\} \delta\left( s-\widetilde{m}_c^2\right)\nonumber\\
&&-\frac{\langle\bar{q} q\rangle^2}{648 T^2}\langle\frac{\alpha_sGG}{\pi}\rangle\int_0^1 dy \, s \, \delta \left( s-\widetilde{m}_c^2\right)\nonumber\\
&&+\frac{\langle\bar{q}g_s\sigma Gq\rangle^2}{384\pi^2T^2}\int_0^1 dy \,s\left( 1+\frac{2s}{9T^2}\right)\, \delta \left( s-\widetilde{m}_c^2\right)
\nonumber\\
&&-\frac{m_c^2\langle\bar{q}q\rangle^2}{216 T^6} \langle\frac{\alpha_sGG}{\pi}\rangle\int_0^1 dy   \, s^2 \, \delta\left( s-\widetilde{m}_c^2\right)\, ,
\end{eqnarray}

\begin{eqnarray}
\rho_{Z}(s)&=&\frac{1}{2048\pi^6}\int dydz \, yz(1-y-z)^3\left(s-\overline{m}_c^2\right)^2\left(21s^2-14s\overline{m}_c^2+\overline{m}_c^4 \right)  \nonumber\\
&&-\frac{1}{2048\pi^6}\int dydz \, yz(1-y-z)^2\left(s-\overline{m}_c^2\right)^4  \nonumber\\
&&-\frac{m_c\langle \bar{q}q\rangle}{32\pi^4}\int dydz \, (y+z)(1-y-z)\left(s-\overline{m}_c^2\right)\left(3s-\overline{m}_c^2\right)  \nonumber \\
&&-\frac{m_c^2}{1536\pi^4} \langle\frac{\alpha_s GG}{\pi}\rangle\int dydz \left( \frac{z}{y^2}+\frac{y}{z^2}\right)(1-y-z)^3 \left\{ 4s-\overline{m}_c^2+\frac{2}{3}\,s^2\,\delta\left(s-\overline{m}_c^2\right)\right\} \nonumber\\
&&+\frac{m_c^2}{1536\pi^4} \langle\frac{\alpha_s GG}{\pi}\rangle\int dydz \left( \frac{z}{y^2}+\frac{y}{z^2}\right)(1-y-z)^2 \left(s-\overline{m}_c^2\right) \nonumber\\
&&-\frac{1}{6144\pi^4}\langle\frac{\alpha_s GG}{\pi}\rangle\int dydz \, (y+z)   (1-y-z)^2 \left(35s^2-40s\overline{m}_c^2+9\overline{m}_c^4 \right) \nonumber\\
&&-\frac{1}{3072\pi^4}\langle\frac{\alpha_s GG}{\pi}\rangle\int dydz \, (y+z)   (1-y-z)  \left(s-\overline{m}_c^2 \right)^2 \nonumber\\
&&+\frac{1}{4608\pi^4}\langle\frac{\alpha_s GG}{\pi}\rangle\int dydz \, (y+z)   (1-y-z)^2  \left(15s^2-16s\overline{m}_c^2+3\overline{m}_c^4 \right) \nonumber\\
&&+\frac{1}{27648\pi^4}\langle\frac{\alpha_s GG}{\pi}\rangle\int dydz \,    (1-y-z)^3 \left(25s^2-24s\overline{m}_c^2+3\overline{m}_c^4 \right) \nonumber\\
&&+\frac{1}{13824\pi^4}\langle\frac{\alpha_s GG}{\pi}\rangle\int dydz \, yz   (1-y-z)  \left(25s^2-24s\overline{m}_c^2+3\overline{m}_c^4 \right) \nonumber\\
&&+\frac{1}{9216\pi^4}\langle\frac{\alpha_s GG}{\pi}\rangle\int dydz \,     (1-y-z)^2  \left(s-\overline{m}_c^2 \right)^2 \nonumber\\
&&+\frac{1}{13824\pi^4}\langle\frac{\alpha_s GG}{\pi}\rangle\int dydz \, yz     \left(s-\overline{m}_c^2 \right)\left(13s-5\overline{m}_c^2 \right)  \nonumber\\
&&+\frac{m_c\langle \bar{q}g_s\sigma Gq\rangle}{64\pi^4}\int dydz  \, (y+z) \left(2s-\overline{m}_c^2 \right) \nonumber\\
&&-\frac{m_c\langle \bar{q}g_s\sigma Gq\rangle}{288\pi^4}\int dydz  \, (1-y-z) \left(2s-\overline{m}_c^2 \right) \nonumber\\
&&+\frac{m_c^2\langle\bar{q}q\rangle^2}{12\pi^2}\int dy \nonumber\\
&&+\frac{g_s^2\langle\bar{q}q\rangle^2}{432\pi^4}\int dydz\,yz\left\{ \left(4-\overline{m}_c^2 \right)+ \frac{2}{3}s^2\delta\left(s-\overline{m}_c^2 \right)\right\} \nonumber\\
&&-\frac{g_s^2\langle\bar{q}q\rangle^2}{1296\pi^4}\int dy\,y(1-y)\left(s-\overline{m}_c^2 \right)\nonumber\\
&&-\frac{g_s^2\langle\bar{q}q\rangle^2}{3888\pi^4}\int dydz\,(1-y-z)\left\{ \left(\frac{z}{y}+\frac{y}{z} \right)9\left(2s-\overline{m}_c^2 \right)+\left(\frac{z}{y^2}+\frac{y}{z^2} \right)\right.\nonumber\\
&&\left.m_c^2\left[ 5+4s\,\delta\left(s-\overline{m}_c^2 \right)\right]+(y+z)\left[6\overline{m}_c^2 +2s^2\,\delta\left(s-\overline{m}_c^2 \right)\right]\right\} \nonumber\\
&&+\frac{g_s^2\langle\bar{q}q\rangle^2}{3888\pi^4}\int dydz\,(1-y-z)\left\{  \left(\frac{z}{y}+\frac{y}{z} \right)\left(-3\overline{m}_c^2 \right)+\left(\frac{z}{y^2}+\frac{y}{z^2} \right)\right. \nonumber\\
&&\left.m_c^2\left[ 2-s\,\delta\left(s-\overline{m}_c^2\right)\right]-(y+z)\left[12s-3\overline{m}_c^2 +2s^2\,\delta\left(s-\overline{m}_c^2\right)\right]\right\} \nonumber
\end{eqnarray}
\begin{eqnarray}
&&+\frac{m_c^3\langle\bar{q}q\rangle}{288\pi^2}\langle\frac{\alpha_sGG}{\pi}\rangle\int dydz \left(\frac{y}{z^3}+\frac{z}{y^3}+\frac{1}{y^2}+\frac{1}{z^2}\right)(1-y-z)\, \delta\left(s-\overline{m}_c^2\right)\nonumber\\
&&-\frac{m_c\langle\bar{q}q\rangle}{96\pi^2}\langle\frac{\alpha_sGG}{\pi}\rangle\int dydz \left(\frac{y}{z^2}+\frac{z}{y^2}\right)(1-y-z) \left\{1+s\,\delta\left(s-\overline{m}_c^2 \right)\right\}\nonumber\\
&&+\frac{m_c\langle\bar{q}q\rangle}{96\pi^2}\langle\frac{\alpha_sGG}{\pi}\rangle\int dydz \left\{1+\frac{1}{3}\,s\,\delta\left(s-\overline{m}_c^2 \right)\right\} \nonumber\\
&&-\frac{m_c\langle\bar{q}q\rangle}{864\pi^2}\langle\frac{\alpha_sGG}{\pi}\rangle\int dydz \left\{\frac{y}{z}+\frac{z}{y}+\left(\frac{1}{y}+\frac{1}{z}\right)(1-y-z)\right\}\left\{1+s\,\delta\left(s-\overline{m}_c^2 \right)\right\} \nonumber\\
&&-\frac{m_c\langle\bar{q}q\rangle}{576\pi^2}\langle\frac{\alpha_sGG}{\pi}\rangle\int dy \left\{1+s\,\delta\left(s-\widetilde{m}_c^2 \right)\right\}\nonumber\\
&&-\frac{m_c^2\langle\bar{q}q\rangle\langle\bar{q}g_s\sigma Gq\rangle}{24\pi^2}\int_0^1 dy \left(1+\frac{s}{T^2} \right)\delta\left(s-\widetilde{m}_c^2\right)\nonumber\\
&&+\frac{ \langle\bar{q}q\rangle\langle\bar{q}g_s\sigma Gq\rangle}{432\pi^2}\int_0^1 dy \,s \, \delta\left(s-\widetilde{m}_c^2\right) \nonumber\\
&&-\frac{m_c^4\langle\bar{q}q\rangle^2}{216T^4}\langle\frac{\alpha_sGG}{\pi}\rangle\int_0^1 dy  \left\{ \frac{1}{y^3}+\frac{1}{(1-y)^3}\right\} \delta\left( s-\widetilde{m}_c^2\right)\nonumber\\
&&+\frac{m_c^2\langle\bar{q}q\rangle^2}{72T^2}\langle\frac{\alpha_sGG}{\pi}\rangle\int_0^1 dy  \left\{ \frac{1}{y^2}+\frac{1}{(1-y)^2}\right\} \delta\left( s-\widetilde{m}_c^2\right)\nonumber\\
&&+\frac{ \langle\bar{q} q\rangle^2}{648 T^2}\langle\frac{\alpha_sGG}{\pi}\rangle\int_0^1 dy \, s \, \delta \left( s-\widetilde{m}_c^2\right)\nonumber\\
&&-\frac{\langle\bar{q}g_s\sigma Gq\rangle^2}{384\pi^2T^2}\int_0^1 dy \,s\left( 1+\frac{2s}{9T^2}\right)\, \delta \left( s-\widetilde{m}_c^2\right)
\nonumber\\
&&+\frac{m_c^2\langle\bar{q}g_s\sigma Gq\rangle^2}{192\pi^2T^6}\int_0^1 dy \, s^2 \, \delta \left( s-\widetilde{m}_c^2\right)
\nonumber\\
&&+\frac{m_c^2\langle\bar{q}q\rangle^2}{216 T^6} \langle\frac{\alpha_sGG}{\pi}\rangle\int_0^1 dy \,   s^2 \, \delta\left( s-\widetilde{m}_c^2\right)\, ,
\end{eqnarray}
 where $\int dydz=\int_{y_i}^{y_f}dy \int_{z_i}^{1-y}dz$, $y_{f}=\frac{1+\sqrt{1-4m_c^2/s}}{2}$,
$y_{i}=\frac{1-\sqrt{1-4m_c^2/s}}{2}$, $z_{i}=\frac{y
m_c^2}{y s -m_c^2}$, $\overline{m}_c^2=\frac{(y+z)m_c^2}{yz}$,
$ \widetilde{m}_c^2=\frac{m_c^2}{y(1-y)}$, $\int_{y_i}^{y_f}dy \to \int_{0}^{1}dy$, $\int_{z_i}^{1-y}dz \to \int_{0}^{1-y}dz$,  when the $\delta$ functions $\delta\left(s-\overline{m}_c^2\right)$ and $\delta\left(s-\widetilde{m}_c^2\right)$ appear.

\section*{Acknowledgements}
This  work is supported by National Natural Science Foundation,
Grant Numbers 11375063, and Natural Science Foundation of Hebei province, Grant Number A2014502017.

\end{document}